\documentclass[%
 reprint,
 amsmath,amssymb,
 aps,onecolumn
]{revtex4-1}

\usepackage{graphicx}% Include figure files
\usepackage{dcolumn}% Align table columns on decimal point
\usepackage{bm}% bold math
\usepackage{braket}

\def\bcen{\begin{center}}
\def\ecen{\end{center}}
\usepackage{epstopdf}
\begin{document}

\preprint{APS/123-QED}

\title{Atomic switch for control of heat transfer in coupled cavities}%Quantum gates in coupled oscillators}% Force line breaks with \\

\author{Nilakantha Meher$^a$}
\author{S. Sivakumar$^b$}%
 \email{palakadsiva@gmail.com}
\affiliation{
 $^a$Department of Physics, Indian Institute of Technology Kanpur, Kanpur, UP 208016, India.
\\
 $^b$ Division of Natural Sciences, Krea University, Andhra Pradesh 517646, India.
}%

%\affiliation{
%\affiliation{
%\affiliation{%



\begin{abstract}
Controlled heat transfer and thermal rectification in a system of two coupled cavities connected to thermal reservoirs are discussed. Embedding a dispersively interacting two-level atom in one of the cavities allows switching from a thermally conducting to resisting behavior. By properly tuning the atomic state and system-reservoir parameters, direction of current can be reversed. It is shown that a large thermal rectification is achievable in this system by tuning the cavity-reservoir and cavity-atom couplings. Partial recovery of diffusive heat transport in an array of $N$ cavities containing one dispersively coupled atom is discussed.     
%\begin{description}
%\end{description}
\end{abstract}

\maketitle

%\tableofcontents

\section{Introduction}
 Coherent and controlled transfer of photons is of fundamental interest in quantum information processing and communication \cite{Northup}. Recent developments in fabrication of suitably coupled cavities have made it possible to study their use in transferring information using photons as the carrier \cite{Meher, Zhong, Yang}. Highly tunable cavity couplings and resonance frequencies of coupled cavities make them suitable for photon transfer, quantum state transfer, entanglement generation, etc \cite{Majumdar, Almeida, Biella,YangLiu,Liao, Liew}. Transport of photons in an array can be modified by embedding atoms or Kerr-medium in the cavities which modify the cavity resonance frequencies \cite{Imamoglu,Felicetti,Qin, Zhou, Brune}. This helps to realize phenomenon such as photon blockade \cite{Imamoglu}, quantum state switching \cite{Meher}, generation of cat states \cite{Brune2}, localization and delocalization \cite{Schmidt, Meher2}, etc.\\

In the ideal case of a cavity being completely isolated from its surroundings, its dynamics is unitary. However, complete isolation of a system is not feasible. In this case, the evolution is not unitary. Simplest of this situation corresponds to coupling the system to a reservoir at zero absolute temperature. On incorporating such a reservoir, many of the photon transport phenomena indicated previously can be explained.  However, the dynamics differs if the cavities are coupled to heat reservoirs at non-zero temperatures, where cavities exchange energy with reservoirs. In this context, coupled cavities can be used to transport energy between the thermal reservoirs \cite{Manzano2,Andr}. For a conventional bulk material, steady state heat transport is governed by  

%A system coupled with two or more thermal reservoirs of different temperatures may reach a non-equilibrium steady state. Continuously  heat energy flows from high temperature to low temperature through the system. Linear response theory works if the system is close to its equilibrium steady state and the system obeys Fourier's law \cite{Garrido, Dhar, Saito2, Michel, Dubi}. According to Fourier's law, the current $(\textbf{J})$ through the system is proportional to the temperature gradient $(\nabla T)$,
\begin{align}\label{Fourier}
\textbf{J}=-\kappa\nabla T,
\end{align}
which is the Fourier's law of heat conduction. Here $\textbf{J}$ is the thermal current and $\nabla T$ is the temperature gradient. The proportionality constant $\kappa$ is the thermal conductivity, which is positive for all known materials. This law is valid if the system is close to its equilibrium, in which case linear response theory is applicable \cite{Garrido, Dhar, Saito2, Michel, Dubi}. Transport of heat by magnetic excitations in spin chains \cite{Saito, Landi, Schachenmayer, Werlang2, Manzano}, phonons in atomic lattices \cite{Thingna, He}, photons in cavity arrays  \cite{Manzano2, Asad,  Purkayastha}, etc.  have been investigated. Similar to electronic devices, several thermal devices such as thermal diodes \cite{Li}, thermal transistors \cite{Li3}, thermal ratchet \cite{NLi}, thermal logic gates \cite{Li4}, thermal memory \cite{Wang},  etc. based on non-equilibrium dynamics have been proposed.\\

 A system away from equilibrium may violate the Fourier's empirical law.   There is no universal theory of heat transfer applicable to all nonequilibrium systems. A  chain of coupled oscillators is known to violate the Fourier's law of heat conduction in the sense that the thermal current is independent of system size and the heat transport is ballistic \cite{Reider, Dhar, Asad, Zotos}. Diffusive transport can be recovered by including anharmonicity or dephasing \cite{Hu,Kamanashish, Asad,Tejal}. The dynamics of nonequilibrium systems is conceptually rich with many unsolved problems.  \\ 

Another interesting phenomenon is thermal rectification, which is essential for realizing thermal diodes and transistors \cite{Starr, Li}. A system shows thermal rectification if it possesses structural asymmetry allowing higher thermal current in one direction. Thermal rectification is known in the case of nanotubes \cite{Chang}, quantum spin chains \cite{Zhang,Yan, Werlang}, nonlinear oscillators \cite{Terr}, two-level systems \cite{Segal}, etc.\\

In the present work, heat transfer in a system of two coupled cavities containing a single atom is discussed. The system-reservoir interaction is assumed to be of Lindblad type \cite{Lindblad}. Magnitude as well as direction of current can be controlled by suitably choosing the atomic state and the system-reservoir parameters. The system exhibits large thermal rectification for proper choices of the cavity-reservoir and cavity-atom couplings.\\ 

The present paper is organized as follows. In Sec. \ref{Physical}, details of the system and its theoretical model are discussed to arrive at an expression for heat current. Also, various special cases of importance are indicated. Based on the dependence of the current on the reservoir temperatures and coupling parameters, violation of Fourier's law is estabslished Sec. \ref{NTC}. Thermal rectification behavior of the system is explored in Sec. \ref{RectificationSection}. Generalization to $N$ cavities is discussed in Sec. \ref{NcavitySection}. Results are summarized in Sec. \ref{Summary}.      
 
% Coupled cavity offers a wide ranges of application in quantum transport \cite{Meher}, entanglement generation, phase transition \cite{Hartmann,Hartmann2}, etc.  
%Controlling photon transport in an array of cavities mediated by a two level atom is discussed earlier \cite{Wang2,Wei,Jing}. In this article we investigate the heat transport properties in a system consist of two coupled cavities containing a single two level atom in anyone of the cavities.  The two cavities are connected to two bosonic reservoirs. 

%The presence of the atom control the current considerably and acts like a thermal switch \cite{Manzano1}. This simple system exhibits negative thermal conductivity  if the system parameter is properly chosen \cite{Iaco}.  Thermal rectification arises due to the asymmetry in the system which is important for thermal diode \cite{Zhang}.    
\section{Current in coupled cavities}\label{Physical}
 A system of two linearly coupled cavities is described by the Hamiltonian\cite{AgarwalBook},
\begin{align}
H_{1}=&\omega_L a^\dagger_L a_L+\omega_R a^\dagger_R a_R+J(a_L^\dagger a_R+a_L a_R^\dagger),
\end{align}
where $\omega_L$ and $\omega_R$ are the resonance frequencies.  The coupling strength between the cavities is  $J$.   In addition, a two-level atom is dispersively coupled to the right cavity and the corresponding atom-cavity interaction is governed by the Hamiltonian \cite{Gerry},
\begin{align}
H_{2}=&\frac{\omega_0}{2}\sigma_z+\chi(\sigma_+\sigma_-+a^\dagger_R a_R\sigma_z),
\end{align}
where $\chi=g^2/(\omega_0-\omega_R)$ is assumed to be positive. The states $\ket{e}$ and $\ket{g}$ are respectively the excited  and ground states of the two-level atom. The operators $\sigma_+=\ket{e}\bra{g}$ and $\sigma_-=\ket{g}\bra{e}$ are the raising and lowering operators for the atom respectively. The energy operator for the atom is $\sigma_z=\ket{e}\bra{e}-\ket{g}\bra{g}$. The coupling strength between the atom and the cavity field is $g$ and the atomic transition frequency is $\omega_0$. This is an effective interaction obtained from Jaynes-Cummings model, if the atom and the cavity are highly detuned so that  $\Delta=(\omega_0-\omega_R) >>g$ and the mean number of photons $n$ is smaller than $\Delta^2/g^2$  \cite{Gerry, Holland}.  Dispersive coupling between atom and cavity has been used to realize the cat states of the cavity field \cite{Brune, Brune2}.\\

The system considered in this work is a pair of linearly  coupled cavities and a dispersively interacting atom in one of the cavities.  Based on the discussion given above, the total Hamiltonian is
\begin{align}\label{Hamiltonian}
H=H_1+H_2=&\frac{\omega_0}{2}\sigma_z+\omega_L a^\dagger_L a_L+\omega_R a^\dagger_R a_R
+\chi(\sigma_+\sigma_-+a^\dagger_R a_R\sigma_z)+J(a_L^\dagger a_R+a_L a_R^\dagger).
\end{align}
This  Hamiltonian conserves the respective total excitation numbers for the cavity fields and the atom in the absence of dissipation,\textit{ i.e.}, $[a_L^\dagger a_L+a_R^\dagger a_R,H]=0$ and $[\sigma_z,H]=0$. As a consequence,  the atom and the field cannot exchange energy in the dispersive limit \cite{Brune}.\\

\begin{figure}[h!]
\centering
\includegraphics[width=8cm,height=4.3cm]{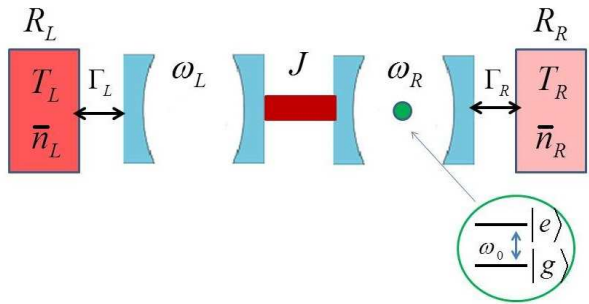}
\caption{Schematic representation of system of coupled cavities with a two-level atom embedded in the right cavity. Both the cavities are also coupled with their respective reservoirs.}
\label{CoupledCavity}
\end{figure}

The system is coupled to two reservoirs, each modelled as a collection of independent oscillators \cite{Carmichael}. The reservoir Hamiltonian is taken to be
\begin{align*}
H_{x}=\sum_{j}\omega_{xj} b_{xj}^\dagger b_{xj}, 
\end{align*}
where $x=L$, $R$ is the index referring to the left reservoir and the right reservoir respectively. The creation and annihilation operators of the reservoirs obey the bosonic commutation relation $[b_{xj}, b^\dagger_{xk}]=\delta_{jk}$. The arrangement of the cavities and reservoirs is shown in Fig. \ref{CoupledCavity}. The interaction Hamiltonian for the cavity-reservoir component is
\begin{align*}
H_{I}=\left(\sum_{j}g_{Lj}\right.&(a_L^\dagger+a_L)(b_{Lj}+b_{Lj}^\dagger)\\
&\left.+\sum_{j}g_{Rj}(a_R^\dagger+a_R)(b_{Rj}+b_{Rj}^\dagger)\right),
\end{align*}
where $g_{Lj}(g_{Rj})$ is the coupling strength of left (right) cavity to $j$th mode of left (right) reservoir.\\  

Under the Born-Markov and rotating wave approximations \cite{GardinerZoller,Zoller}, the reduced joint density matrix for the two cavities (traced over the reservoirs) obeys \cite{Carmichael}
\begin{align}\label{Master}
\frac{\partial \rho}{\partial t}=-i[H,\rho]+\mathcal{D}_L(\rho)+\mathcal{D}_R(\rho),
\end{align}
where the Lindblad operators
\begin{align}\label{Lindblad}
\mathcal{D}_x(\rho)=&\frac{\Gamma_x(\bar n_x+1)}{2}(2a_x\rho a_x^\dagger-a_x^\dagger a_x \rho-\rho a_x^\dagger a_x)\nonumber\\
&+\frac{\Gamma_x \bar n_x}{2}(2a_x^\dagger\rho a_x-a_x a_x^\dagger \rho-\rho a_x a_x^\dagger),
\end{align}
for $x=L,R$. The parameters $\Gamma_L$ and $\Gamma_R$ are related to the coupling strengths as \cite{Biehs}
\begin{align}
\Gamma_x=2\pi\sum_{j}g_{xj}^2\delta(\omega_{xj}-\omega_x).
\end{align}
 The two terms in Eqn. \ref{Lindblad} correspond to energy flow from the system to the reservoir and vice-versa respectively. The dynamics generated by the master equation approach satisfies the detailed balance condition and gives the correct steady state if the different components of the system are weakly coupled \cite{Rivas, Purkayastha2,Manzano, Santos}.\\
 
The reservoirs $R_L$ and $R_R$ are assumed to be in thermal equilibrium at temperatures $T_L$ and $T_R$ respectively. The density operators which characterize the states of the reservoirs are
\begin{align}
\varrho_x=\frac{e^{-H_x/k_BT_x}}{\text{Tr}\left(e^{-H_x/k_BT_x}\right)},
\end{align} 
 with mean photon numbers 
\begin{align}
\bar n_{x}=\frac{1}{\exp{(\omega_{x}/k_B T_x)}-1},
\end{align} 
where $x=L,R$.\\

The dynamics of the system can be understood from the temporal evolution of expectation values of various operators. The expectation values satisfy  
\begin{widetext}
\begin{subequations}\label{EquationOfMotion}
\begin{eqnarray}
&\frac{d}{dt}\langle a_L^\dagger a_L\rangle=iJ(\langle a_L a_R^\dagger\rangle-\langle a_L^\dagger a_R\rangle)-\Gamma_L \langle a_L^\dagger a_L\rangle+\Gamma_L \bar{n}_L,\\
&\frac{d}{dt}\langle a_R^\dagger a_R\rangle=-iJ(\langle a_L a_R^\dagger\rangle-\langle a_L^\dagger a_R\rangle)-\Gamma_R \langle a_R^\dagger a_R\rangle+\Gamma_R \bar{n}_R,\\
&\frac{d}{dt}\langle a_L^\dagger a_R\rangle=i\Delta_c\langle a_L^\dagger a_R\rangle-iJ(\langle a_L^\dagger a_L\rangle-\langle a_R^\dagger a_R\rangle)-i\chi\langle a_L^\dagger a_R\sigma_z\rangle-\gamma\langle a_L^\dagger a_R\rangle,\\ 
&\frac{d}{dt}\langle a_L a_R^\dagger\rangle=-i\Delta_c\langle a_L a_R^\dagger\rangle+iJ(\langle a_L^\dagger a_L\rangle-\langle a_R^\dagger a_R\rangle)+i\chi\langle a_L a_R^\dagger\sigma_z\rangle-\gamma\langle a_L a_R^\dagger \rangle,\\
&\frac{d}{dt}\langle a_L^\dagger a_L\sigma_z\rangle=iJ(\langle a_L a_R^\dagger \sigma_z\rangle-\langle a_L^\dagger a_R \sigma_z\rangle)-\Gamma_L \langle a_L^\dagger a_L \sigma_z\rangle+\Gamma_L \bar{n}_L\langle \sigma_z\rangle,\\
&\frac{d}{dt}\langle a_R^\dagger a_R\sigma_z\rangle=-iJ(\langle a_L a_R^\dagger\sigma_z\rangle-\langle a_L^\dagger a_R\sigma_z\rangle)-\Gamma_R \langle a_R^\dagger a_R\sigma_z\rangle+\Gamma_R \bar{n}_R\langle \sigma_z\rangle,\\
&\frac{d}{dt}\langle a_L^\dagger a_R\sigma_z\rangle=i\Delta_c\langle a_L^\dagger a_R\sigma_z\rangle-iJ(\langle a_L^\dagger a_L\sigma_z\rangle-\langle a_R^\dagger a_R\sigma_z\rangle)-i\chi\langle a_L^\dagger a_R\rangle-\gamma\langle a_L^\dagger a_R\sigma_z\rangle,\\ 
&\frac{d}{dt}\langle a_L a_R^\dagger\sigma_z\rangle=-i\Delta_c\langle a_L a_R^\dagger\sigma_z\rangle+iJ(\langle a_L^\dagger a_L\sigma_z\rangle-\langle a_R^\dagger a_R\sigma_z\rangle)+i\chi\langle a_L a_R^\dagger\rangle-\gamma\langle a_L a_R^\dagger\sigma_z \rangle,
\end{eqnarray}
\end{subequations}
where $\gamma=(\Gamma_L+\Gamma_R)/2$ and $\Delta_c=\omega_L-\omega_R$. Here $\langle A \rangle=\text{Tr}[\rho A]$, where $\rho$ satisfies the master equation given in Eqn. \ref{Master}. \\
\end{widetext}

As $[\sigma_z, H]=[\sigma_z,\mathcal{D}_L(\rho) ]=[\sigma_z, \mathcal{D}_R(\rho)]=0$, the evolution equation for $\langle\sigma_z\rangle$ is $d \langle \sigma_z\rangle/d t=0$. This indicates that the value of $\langle \sigma_z \rangle $ remains constant during time evolution as a consequence of the fact that the atom is dispersively coupled with the cavity field.\\

Steady state  current is defined \textit{via} the continuity equation 
\begin{align}\label{Cont}
\frac{d}{dt}\langle H\rangle=0,
\end{align}
which expresses the conservation of the total energy in the system.\\
  
With $\langle H \rangle=$Tr$[\rho H]$ and using Eqn. \ref{Master} for evolving $\rho$, the continuity equation given in Eqn. \ref{Cont} yields
\begin{align}
0=\text{Tr}[H\mathcal{D}_L(\rho)+H\mathcal{D}_R(\rho)]=:I_L+I_R.
\end{align} 
Here $I_x=\text{Tr}[H\mathcal{D}_x(\rho)], x=L,R$. Further, $I_L$ refers to the thermal current from the left reservoir $R_L$ to the system and $I_R$ indicates the current from  the right reservoir $R_R$ to the system. Using Eqn. \ref{Master}, the steady state heat current from the left reservoir to the right reservoir through the system is 
\begin{align}\label{CurrentExpression}
I_L=\text{Tr}[H\mathcal{D}_L(\rho)]=\Gamma_L(I_{nd}-I_{coh}).
\end{align}
%\begin{align}
%I_L=\Gamma_L(\bar{n}_L-\langle a_L^\dagger a_L \rangle_{ss})\omega_L-\frac{\Gamma_L}{2}J(\langle a_L^\dagger a_R\rangle_{ss}+\langle a_L a_R^\dagger \rangle_{ss}).
%\end{align}
%It is to be noted that, the current $I_L$ can be written as
Here $I_{nd}=(\bar{n}_L-\langle a_L^\dagger a_L \rangle_{ss})\omega_L$ is the current due to mean excitation number difference between the left reservoir and the left cavity, and $I_{coh}=\frac{1}{2}J(\langle a_L^\dagger a_R\rangle_{ss}+\langle a_L a_R^\dagger \rangle_{ss})$ is the current due to the total coherence in the system. Here $\langle \cdot \rangle_{ss}$ represents the steady state mean value.  A similar expression for the steady state heat current from the right reservoir to the left reservoir is
\begin{align}\label{RightCurrent}
I_R&=\text{Tr}[H\mathcal{D}_R(\rho)],\nonumber\\
&=\Gamma_R(\bar{n}_R-\langle a_R^\dagger a_R \rangle_{ss})(\omega_R+\langle\sigma_z\rangle\chi)-\Gamma_R I_{coh}.
\end{align}\\
Steady state solutions are obtained by equating the time derivatives of the expectation values of the relevant operators given in Eqns. \ref{EquationOfMotion}$a$-\ref{EquationOfMotion}$h$ to zero. The steady state values are 
\begin{subequations}\label{SteadyMean}
\begin{eqnarray}
&\langle a_L^\dagger a_L \rangle_{ss}=\frac{C(\Gamma_L \bar{n}_L+\Gamma_R \bar{n}_R)+\Gamma_L\Gamma_R \bar{n}_L}{C(\Gamma_L+\Gamma_R)+\Gamma_L\Gamma_R},\\\nonumber\\
&\langle a_R^\dagger a_R \rangle_{ss}=\frac{C(\Gamma_L \bar{n}_L+\Gamma_R \bar{n}_R)+\Gamma_L\Gamma_R \bar{n}_R}{C(\Gamma_L+\Gamma_R)+\Gamma_L\Gamma_R},\\\nonumber\\
&\delta N= \langle a_L^\dagger a_L \rangle_{ss}-\langle a_R^\dagger a_R \rangle_{ss}=\frac{\Gamma_L\Gamma_R(\bar{n}_L-\bar{n}_R)}{C(\Gamma_L+\Gamma_R)+\Gamma_L\Gamma_R},\\\nonumber\\
&\text{and}~~~~~~~~~~~~~~~~~~~~~~~~~~~~~~~~~~~~~~~~~~~~~~~~~~~~~~~~~~~~~~~~~~~~~\nonumber\\
&\langle a_L^\dagger a_R \rangle_{ss}=-J\frac{\chi \langle\sigma_z\rangle+\Delta_c+i\gamma}{\chi^2-\Delta_c^2+\gamma^2-2i\gamma\Delta_c}\delta N,
\end{eqnarray}
\end{subequations}
with 
\begin{align}
C=2J^2\gamma\frac{\Delta_c^2+\chi^2+2\Delta_c\chi\langle \sigma_z\rangle+\gamma^2}{(\chi^2-\Delta_c^2+\gamma^2)^2+4\gamma^2\Delta_c^2}.
\end{align} \\ 

Using these steady state solutions, Eqn. \ref{CurrentExpression} yields
\begin{widetext}
\begin{align}\label{nonresonantCurrent}
I_L=J^2\delta N\frac{\Gamma_L\chi\langle\sigma_z\rangle(\chi^2-\Delta_c^2+\gamma^2)+(\omega_L\Gamma_R+\omega_R\Gamma_L)(\Delta_c^2+\gamma^2)+\chi^2(2\omega_L\gamma+\Delta_c\Gamma_L)+4\Delta_c\chi\langle\sigma_z\rangle \omega_L\gamma}{(\chi^2-\Delta_c^2+\gamma^2)^2+4\gamma^2\Delta_c^2}.
\end{align}
%I_L=J^2\delta N\frac{2\omega_L\gamma(\Delta_c^2+\chi^2+\gamma^2)+\Gamma_L\chi\langle\sigma_z\rangle(\chi^2-\Delta_c^2+\gamma^2)+\Gamma_L\Delta_c(\chi^2-\Delta_c^2-\gamma^2)+4\Delta_c\chi\langle\sigma_z\rangle \omega_L\gamma}{(\chi^2-\Delta_c^2+\gamma^2)^2+4\gamma^2\Delta_c^2}.
\end{widetext}
In the absence of inter-cavity coupling $(J=0)$, the cavities equilibrate with their respective reservoirs with mean photon numbers $\bar{n}_L$ and $\bar{n}_R$. The currents $I_L$ and $I_R$ vanish since energy cannot flow from one cavity to another as $J=0$. If the coupling is non-zero and the reservoirs are at different temperatures, energy flows from one reservoir to other through the cavities.\\

Interestingly, expression in Eqn. \ref{nonresonantCurrent} shows that the current through the system explicitly depends on $\langle\sigma_z\rangle$, which, in turn, depends on the state of the atom. This dependency arises as the atom modifies the cavity resonance frequency and the coherences $\langle a_L^\dagger a_R \rangle_{ss}$ and $\langle a_L a_R^\dagger \rangle_{ss}$, as well.  By a proper choice of the atomic state, $\langle\sigma_z\rangle$ can be tuned from $+1$ corresponding to the atom in its excited state to $-1$, \textit{i.e.}, the atom is in its ground state. This feature can be used to control the energy flow (current) between the reservoirs. \\

%In the absence of atom, the current through the coupled cavities is

%\begin{align}
%J_{c}=J^2\delta N\frac{\omega_L\Gamma_R+\omega_R\Gamma_L}{(\gamma^2-\Delta_c^2)^2+4\gamma^2\Delta_c^2}(\gamma^2+\Delta_c^2).
%\end{align}

If the cavities are resonant, \textit{i.e.}, $\omega_L=\omega_R=\omega $, equivalently, $\Delta_c=0$. In the absence of the atom, the total coherence is zero $\langle a_L^\dagger a_R \rangle_{ss}+\langle a_L a_R^\dagger \rangle_{ss}=0,$ as can be seen from Eqn. \ref{SteadyMean}$d$.  
%The current in the system is result of the mean photon number difference between one of the reservoirs and a cavity, \textit{i.e.}, $\bar{n}_L-\langle a_L^\dagger a_L \rangle_{ss}$. 
The current through the cavities is
\begin{align}\label{ResonantCurrentAbsentAtom}
I_L=\frac{4\omega J^2\Gamma_L\Gamma_R}{(4J^2+\Gamma_L\Gamma_R)(\Gamma_L+\Gamma_R)}(\bar{n}_L-\bar{n}_R).
\end{align}
The current is proportional to the difference in the mean photon numbers; equivalently, the current is proportional to the temperature difference of the two reservoirs for a fixed system size, which is like the Fourier's law.\\
% This is the expression is obtained for current in an array of $N$ linearly-coupled resonant cavities as well. The current doesn't depend on the size of the array \cite{Asad}. This violates the Fourier's law which implies an inverse relationship between the current and the system size. \\

 If the temperatures of the two reservoirs are equal ($\bar{n}_L=\bar{n}_R=\bar{n}$), the system equilibrates with the reservoirs and no current flows through the system. The mean number of photons in the cavities are $ \langle a_L^\dagger a_L \rangle_{ss}=\langle a_R^\dagger a_R \rangle_{ss}=\bar{n}$. Also, the states of the cavity fields satisfy the zero coherence condition, namely, $ \langle a_L^\dagger a_R \rangle_{ss}=\langle a_R^\dagger a_L \rangle_{ss}=0$.
To know the states of the fields in the cavities, the fidelity $F(\rho_{th},\rho_x)$
\begin{align}
F(\rho_{th},\rho_x)=\text{Tr}\left(\sqrt{\sqrt{\rho_{th}}\rho_x\sqrt{\rho_{th}}}\right),
\end{align}
between the thermal field and the cavity field is calculated. Here
\begin{align}\label{thermalstate}
\rho_{th}=\frac{1}{1+\bar{n}}\sum_{n=0}^\infty \left(\frac{\bar{n}}{1+\bar{n}}\right)^n \ket{n}\bra{n},
\end{align}
is the single mode Gibbs thermal state; $\rho_x (x=L,R)$ are the steady state reduced density matrices for the left- and right-cavities respectively. The steady state fidelity $F(\rho_{th},\rho_x)$ is unity.  Therefore, the cavity fields are also Gibbs thermal state. The second order correlation function
\begin{align}
g^{(2)}_x(0)=\frac{\text{Tr}(\rho_x a_x^{\dagger 2} a_x^2)}{\left[\text{Tr}(\rho_x a_x^{\dagger} a_x)\right]^2},
\end{align}
in the steady state $\rho_x$ is $2$ same as that of the thermal state. This confirms that the cavity states are thermal states $\rho_{th}$.\\

 If a temperature difference is maintained between the reservoirs, the high temperature reservoir is the source of energy to the system and the low temperature reservoir is the sink for the energy to establish a steady state. As a consequence, heat continuously flows from the high temperature reservoir to the low temperature reservoir. The system reaches a non-equilibrium steady state with effective mean photon numbers  $\langle a_L^\dagger a_L \rangle_{ss}$ and $\langle a_R^\dagger a_R \rangle_{ss}$  in the left- and right- cavities respectively. Analytical expressions for these mean photon numbers are given in Eqn. \ref{SteadyMean}$a$ and Eqn. \ref{SteadyMean}$b$. A non-equilibrium steady state is not necessarily the Gibbs thermal state. \\

In the presence of an atom in one of the cavities,  as shown in Fig. \ref{CoupledCavity}, the current through the system is
\begin{align}\label{Current}
I_L=\Theta\frac{\bar{c}}{\gamma} \left(\gamma\omega+\frac{\Gamma_L}{2}\chi\langle \sigma_z \rangle\right)(\bar{n}_L-\bar{n}_R),
\end{align}
where 
\begin{align*}
\Theta=\frac{\Gamma_L\Gamma_R}{\bar{c}(\Gamma_L+\Gamma_R)+\Gamma_L\Gamma_R},
\end{align*}
and $\bar{c}=2J^2\gamma/(\chi^2+\gamma^2)$.\\

\begin{figure}[h!]
\centering
\includegraphics[width=9cm,height=6cm]{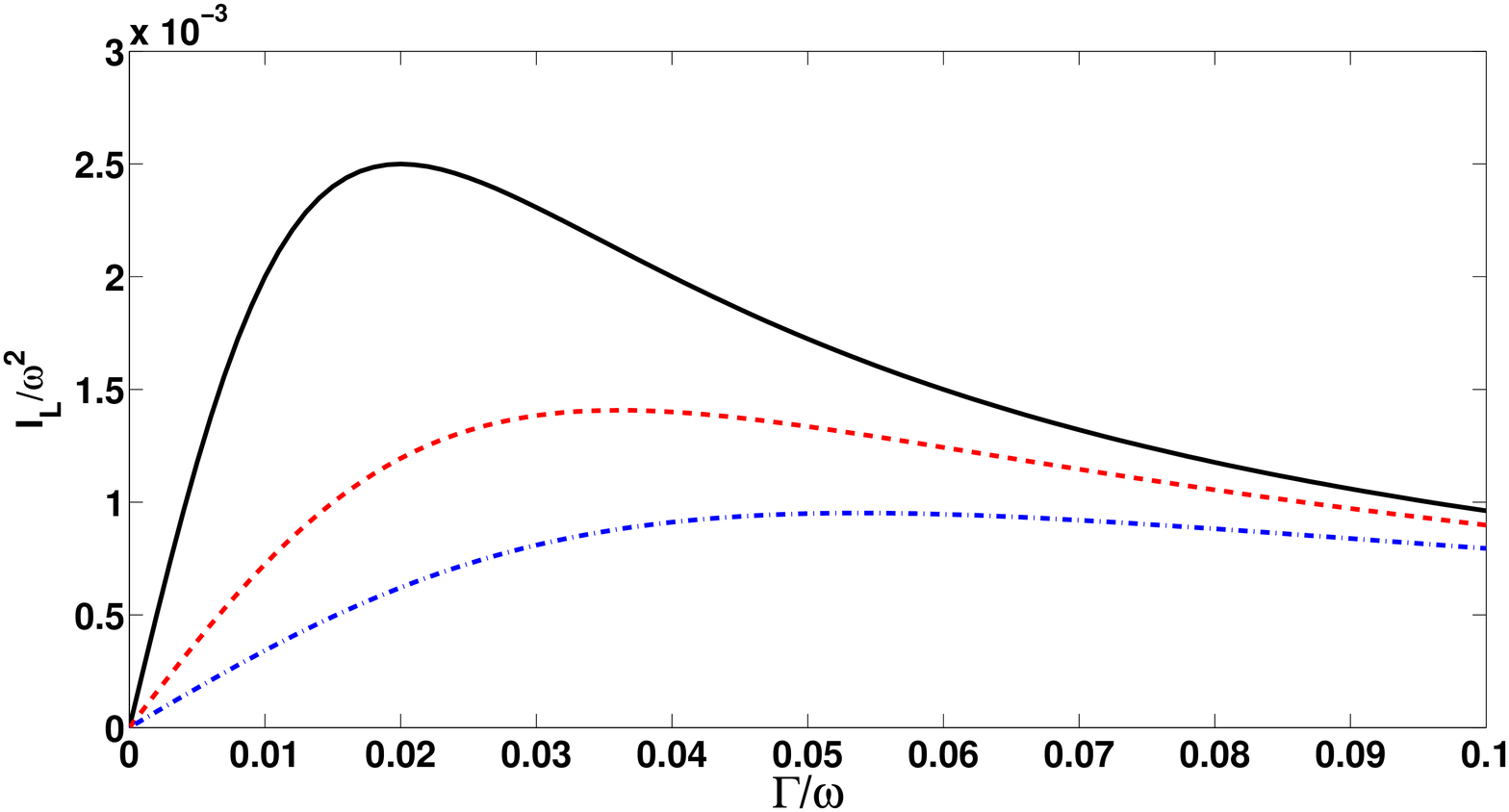}
\caption{Current $I_L/\omega^2$ shown as a function of reservoir coupling strength $\Gamma/\omega$ for different atom-cavity coupling strengths $\chi/\omega=0$ (continuous), $0.03$ (dashed) and $0.05$ (dot-dashed). The system-reservoir parameters are $J/\omega=0.02, \bar{n}_L-\bar{n}_R=0.5 $ and $\langle \sigma_z\rangle=1$.}
\label{Equalbathcouplings}
\end{figure}

If $\Gamma_L=\Gamma_R=\Gamma$, then  
\begin{align}\label{CurrentEqualGamma}
I_L=2J^2\frac{\Gamma}{4J^2+\chi^2+\Gamma^2}\left(\omega+\frac{\chi}{2}\langle\sigma_z\rangle\right)(\bar{n}_L-\bar{n}_R).
\end{align}
Scaled current $I_L/\omega^2$ as a function of $\Gamma/\omega$ is shown in Fig. \ref{Equalbathcouplings}. Maximum current flows through the system if $\Gamma=\sqrt{4J^2+\chi^2}$. This special value $\sqrt{4J^2+\chi^2}$ corresponds to the Rabi frequency of the oscillation of the mean number of photon when the cavity detuning is $\chi$ and the cavities are not coupled to the reservoirs. The detuning between the cavity frequencies arises due to the atom in one of the cavities. The competition between the cavity-reservoir energy exchange rate $\Gamma$ and the cavity-cavity energy exchange rate $\sqrt{4J^2+\chi^2}$ affects the current through the system. If the two rates are equal, then
\begin{align}
I_L=\frac{J^2}{\sqrt{4J^2+\chi^2}}\left(\omega+\frac{\chi}{2}\langle\sigma_z\rangle\right)(\bar{n}_L-\bar{n}_R),
\end{align}
which is the maximum current. If $\Gamma>>\sqrt{4J^2+\chi^2}$, the cavities and their respective reservoirs exchange energy faster than the inter-cavity exchange. In the opposite limit, both the cavities exchange energy with each other faster than with their respective reservoirs. This mismatch between the energy exchange rates reduces the current. From Eqn. \ref{CurrentEqualGamma}, it is seen that for small $\Gamma$, $I_L \propto \Gamma$ and  for large $\Gamma$,  $I_L \propto \Gamma^{-1}$. It may be noted that a system of three cavities containing two 3-level atoms has also been shown to allow control of magnitude of heat current \cite{Manzano2}.

\section{Negative thermal conductivity}\label{NTC}
According to the Fourier's law given in Eqn. \ref{Fourier}, current is proportional to temperature gradient. Using the fact that $\delta N\propto (\bar{n}_L-\bar{n}_R)$ as given in Eqn. \ref{SteadyMean}$c$, the expression for $I_L$ in Eqn. \ref{nonresonantCurrent} can be written in the form
\begin{align}
I_L=\tilde{\kappa}(\bar{n}_L-\bar{n}_R),
\end{align}
for comparing with the Fourier's law.
Here $\tilde{\kappa}$ is the effective thermal conductivity. It is to be noted that thermal conductivity can be tuned by suitably choosing the atomic state.
Two important cases corresponding to the atom being in the excited state and the ground state are considered, \textit{i.e.}, $\langle \sigma_z\rangle=\pm 1$. The corresponding currents are
\begin{align}\label{Currentpm}
I_{L}=J^2\delta N\frac{\Omega}{(\chi^2-\Delta_c^2+\gamma^2)^2+4\gamma^2\Delta_c^2}((\Delta_c \pm\chi)^2+\gamma^2).
\end{align}
where $\Omega=\omega_L\Gamma_R+\Gamma_L(\omega_R\pm\chi)$. We assume $\bar{n}_L> \bar{n}_R$, \textit{i.e.}, $\delta N>0$ for subsequent discussion. In this assumption, $I_L$ and $\Omega$ have the same sign. If the atom is in its ground state, sign of $\Omega$ is changeable by properly choosing the ratios $\Gamma_R/\Gamma_L$ and $(\chi-\omega_R)/\omega_L$. Consequently, direction of current can also be changed. It is to be pointed out that $\omega_R-\chi$ is the resonance frequency of the right cavity modified by the atom. If the atom is in its excited state, \textit{i.e.}, $\langle \sigma_z\rangle=+1$, $I_L$ is always positive, meaning the thermal current flows from the high temperature reservoir to the low temperature reservoir (conventional flow) and reversal of current is not possible. \\

 % This amounts to say that, the direction of current can be switched from conventional current flow (hot to cold) to unconventional current flow (cold to hot) by switching the atomic state. and suitably choosing the ratios $\Gamma_R/\Gamma_L$ and $(\chi-\omega_R)/\omega_L$. It may be pointed out that $\omega_R-\chi$ is the resonance frequency of the right cavity modified due to the atom. \\
In order to exhibit the switching action by the atom, we choose $\chi>\omega_R$. If the system-reservoir parameters satisfy
\begin{align}\label{HightoLow}
\frac{\Gamma_R}{\Gamma_L}> \frac{(\chi-\omega_R)}{\omega_L},
\end{align}
thermal current flows from the high temperature reservoir to the low temperature reservoir, independent of the atomic state.\\

If the ratios are equal, \textit{i.e.}, 
\begin{align}\label{Zerocurrent}
\frac{\Gamma_R}{\Gamma_L}=\frac{(\chi-\omega_R)}{\omega_L},
\end{align}
and the atom is in the ground state, the thermal current through the system is zero even if the reservoirs are at different temperatures. The system completely blocks the heat flow like a thermal insulator. By switching the atom to its excited state, the system changes from a thermal-insulator to a thermal-conductor. \\

If the atom is in its ground state and the system-reservoir parameters are such that
\begin{align}\label{LowtoHigh}
\frac{\Gamma_R}{\Gamma_L}< \frac{(\chi-\omega_R)}{\omega_L},
\end{align}
then $\Omega<0$ and
the direction of thermal current reverses, \textit{i.e.}, current flows from low temperature reservoir to high temperature reservoir (unconventional flow). In such case, thermal conductivity of the system can be interpreted to be negative in which case heat flows from the low temperature to high temperature. Emergence of this negative current may be a result of coupling the system to Markovian baths \cite{Zoller}.
By switching the atom from its ground state to excited state, the unconventional flow of thermal current switches to the conventional flow. Thus, the atom acts as a thermal switch which brings about a controllable current flow through the cavities.\\

 To summarize, we define
\begin{align}
\alpha=\frac{\Gamma_R/\Gamma_L}{(\chi-\omega_R)/\omega_L}.
\end{align}
The three conditions given in Eqns. (\ref{HightoLow}-\ref{LowtoHigh}) correspond to $\alpha$ becoming greater than, equal to or less than unity respectively. The signs of the respective currents established in the system are indicated in Table. \ref{AtomStateCurrent}.\\
\begin{table}
\caption{Conditions for positive and negative thermal currents.}
\begin{center}
\begin{tabular}{ | c | c | c | } 
 \hline
 & $~~\langle \sigma_z\rangle =+1~~$ & $~~\langle \sigma_z\rangle =-1~~$ \\ [1ex]
 \hline
 $~~\alpha>1~~$  & $I_L>0$ & $I_L>0$ \\ [1ex]

 $\alpha=1$ & $I_L>0$ & $I_L=0$ \\  [1ex]

 $\alpha<1$ & $I_L>0$ & $I_L<0$ \\ 
 \hline
\end{tabular}
\label{AtomStateCurrent}
\end{center}
\end{table}

\begin{figure}
\centering
\includegraphics[width=9cm,height=5.5cm]{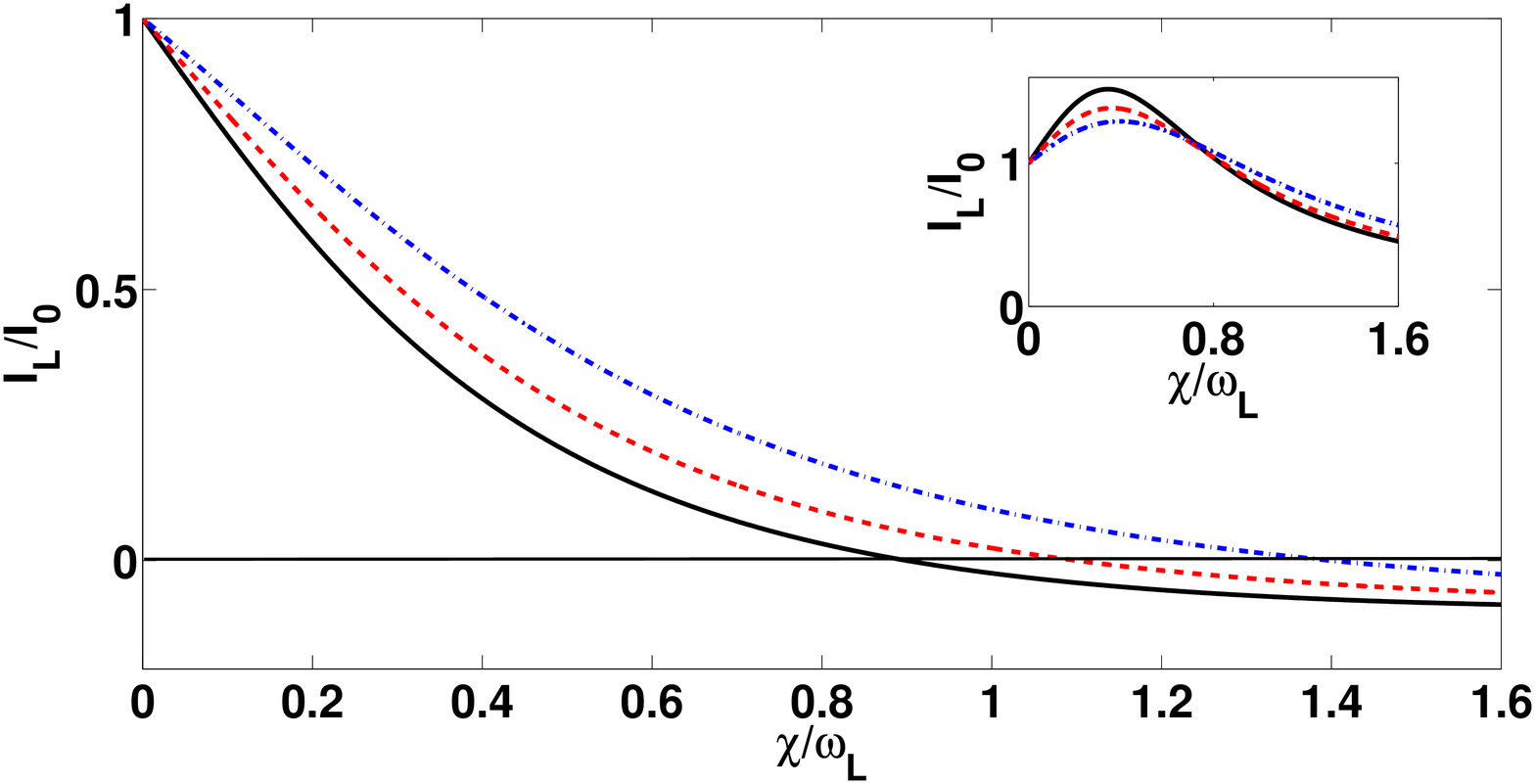}
\caption{Normalized current $I_{L}/I_0$ shown as a function of $\chi/\omega_L$  for different cavity-reservoir coupling ratio $\Gamma_R/\Gamma_L=0.1$ (continuous), $0.3$ (dashed) and $0.6$ (dot-dashed). The system-reservoir parameters are $J/\omega_L=0.05, \bar{n}_L-\bar{n}_R=0.5, \omega_R/\omega_L=0.8$ and $\langle \sigma_z\rangle=-1$. The current in the system is shown in the inset for the same values of the parameters and $\langle \sigma_z\rangle=+1$.}
\label{negativecurrent}
\end{figure}

Scaled current $I_{L}/I_0$ for the case of the atom in its ground state is shown as a function of $\chi/\omega_L$ in Fig. \ref{negativecurrent} for cavity-reservoir coupling ratios $\Gamma_R/\Gamma_L=0.1$ (continuous), $0.3$ (dashed) and $0.6$ (dot-dashed). Here $I_0$ is the amount of current flowing through the system when $\chi=0$. The inset figure shows the scaled current in the system when the atom is in excited state for the same values of $\Gamma_R/\Gamma_L$.  Note that the current is always positive if the atom is in the excited state (inset figure). If the atom is in its ground state, current vanishes if the system and reservoir parameters satisfy Eqn. \ref{Zerocurrent}. Negative current occurs at different values of $\chi/\omega_L$ required to satisfy Eqn. \ref{LowtoHigh}.

\begin{figure}[h!]
\centering
\includegraphics[width=9cm,height=5cm]{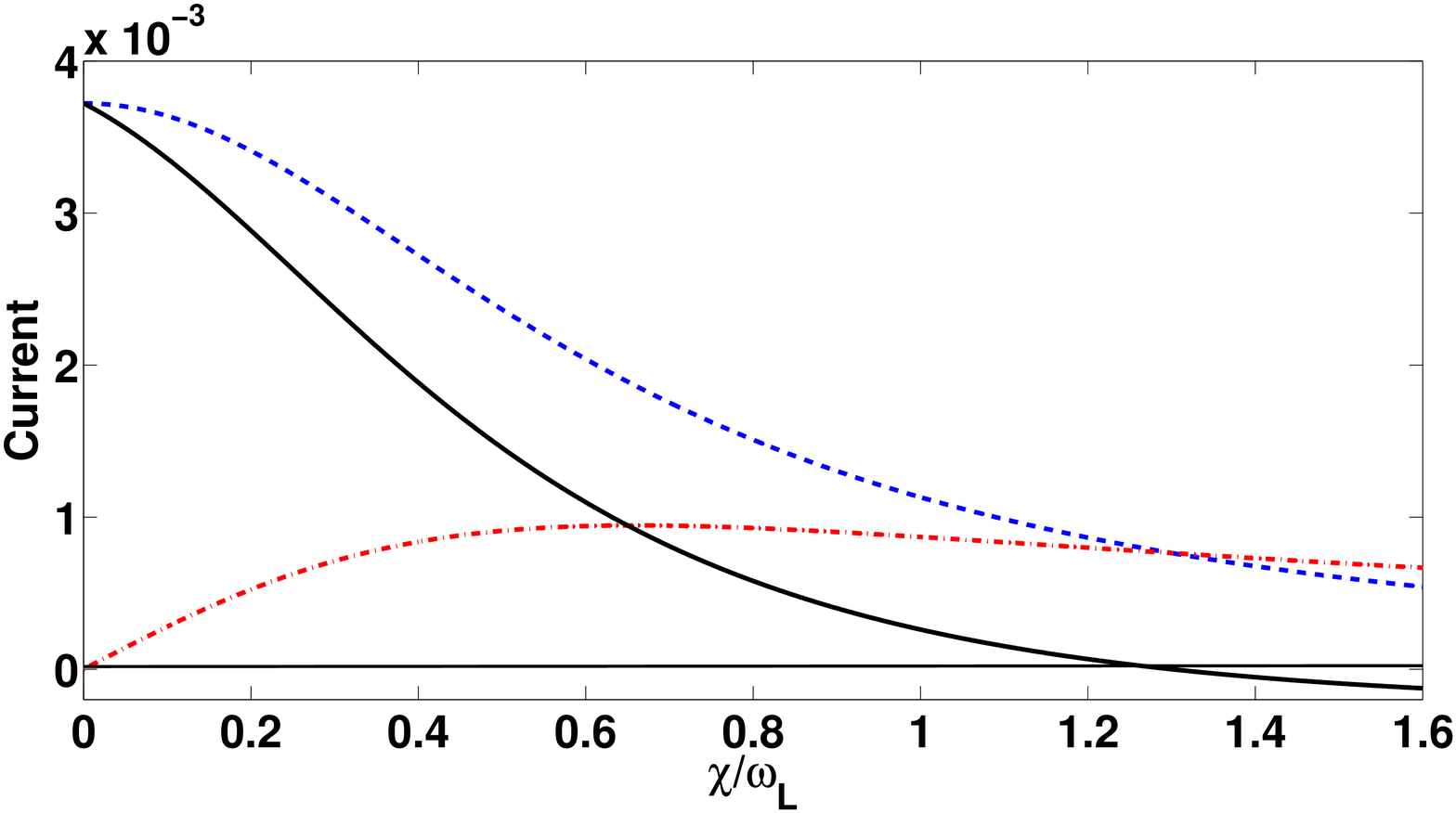}
\caption{Dimensionless currents $I_{L}/\omega_L^2$(continuous), $I_{nd}/\omega_L^2$ (dashed) and $I_{coh}/\omega_L^2$ (dot-dashed) shown as function of $\chi/\omega_L$.  Here  $\Gamma_R/\Gamma_L=0.3$, $J/\omega_L=0.05, \bar{n}_L-\bar{n}_R=0.5, \omega_R/\omega_L=1$ and $\langle \sigma_z\rangle=-1$.}
\label{Currentcontribution}
\end{figure}

Negative current arises because the contribution from the coherence part $I_{coh}$ is more than the current due to mean excitation number difference $I_{nd}$, which makes $I_L$ negative (refer Eqn. \ref{CurrentExpression}). Dimensionless quantities $I_{nd}/\omega_L^2,I_{coh}/\omega_L^2$ and $I_{L}/\omega_L^2$ are shown in Fig. \ref{Currentcontribution} as a function of the atom-field coupling strength $\chi/\omega_L$. It is to be noted that if the parameters are chosen to satisfy  Eqn. \ref{Zerocurrent}, in which case $I_{nd}=I_{coh}$, the system completely blocks the current. Current reverses its direction from the low temperature reservoir to the high temperature reservoir when $I_{coh}>I_{nd}$. In this sense, the coherence in the system drives energy to flow to the high temperature reservoir.    
\\

%\textbf{It is of interest to note that if the Eqn. \ref{LowtoHigh} for negative current is satisfied, the thermal conductivity $\tilde{\kappa}$ is negative. Negative thermal conductivity is also observed in a chain of rotors \cite{Iaco}. It is necessary to mention that the current flows against the natural arrow of heat, which is not possible in the absence of atom. Hence it violates the second law of thermodynamics, if the dispersive coupling does not cost any external energy. } 

\section{Thermal Rectification}\label{RectificationSection}

 \begin{figure}
\centering
\includegraphics[width=8cm,height=4.5cm]{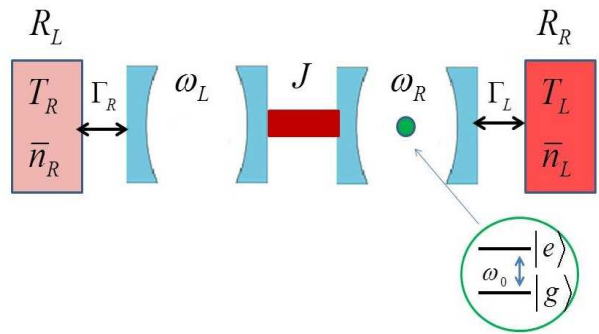}
\caption{Reverse configuration of system-reservoirs. reservoir temperatures and the system-reservoir coupling strengths are interchanged.}
\label{CoupledCavityReverse}
\end{figure}

A system exhibits thermal rectification if thermal current depends on the direction of heat flow,
\begin{align}
I(\Delta n)\neq -I(-\Delta n), 
\end{align}  
where $\Delta n=\bar{n}_L-\bar{n}_R$ is the difference in the average photon number of the left and right reservoirs.
This means that by swapping the thermal reservoirs, current changes both sign and magnitude.\\

If the system is symmetric under the exchange of cavities, rectification is not possible. In the system under discussion, assymmetry is due to presence of the atom in one of the cavities. Thermal rectification is to be established by studying the transport of photon in the reverse configuration  
realized by interchanging the reservoirs and system-reservoir coupling strengths. The reverse configuration is shown in Fig. \ref{CoupledCavityReverse}. The relevant Lindblad operators for the reverse configuration are
\begin{align}
\mathcal{D}_L(\rho)=&\frac{\Gamma_R(\bar n_R+1)}{2}(2a_L\rho a_L^\dagger-a_L^\dagger a_L \rho-\rho a_L^\dagger a_L)\nonumber\\
&+\frac{\Gamma_R \bar n_R}{2}(2a_L^\dagger\rho a_L-a_L a_L^\dagger \rho-\rho a_L a_L^\dagger),\nonumber\\
\mathcal{D}_R(\rho)=&\frac{\Gamma_L(\bar n_L+1)}{2}(2a_R\rho a_R^\dagger-a_R^\dagger a_R \rho-\rho a_R^\dagger a_R)\nonumber\\
&+\frac{\Gamma_L \bar n_L}{2}(2a_R^\dagger\rho a_R-a_R a_R^\dagger \rho-\rho a_R a_R^\dagger).\nonumber
\end{align}
The atom is taken to be in its ground state. Steady state solutions for the expectation values of operators for the reverse configuration can be obtained by the transformations $\Gamma_L\longrightarrow \Gamma_R$, $\bar n_L\longrightarrow \bar n_R$ and vice-versa in Eqns. \ref{SteadyMean}$a$-\ref{SteadyMean}$d$.

\begin{figure}
\centering
\includegraphics[width=9cm,height=5.5cm]{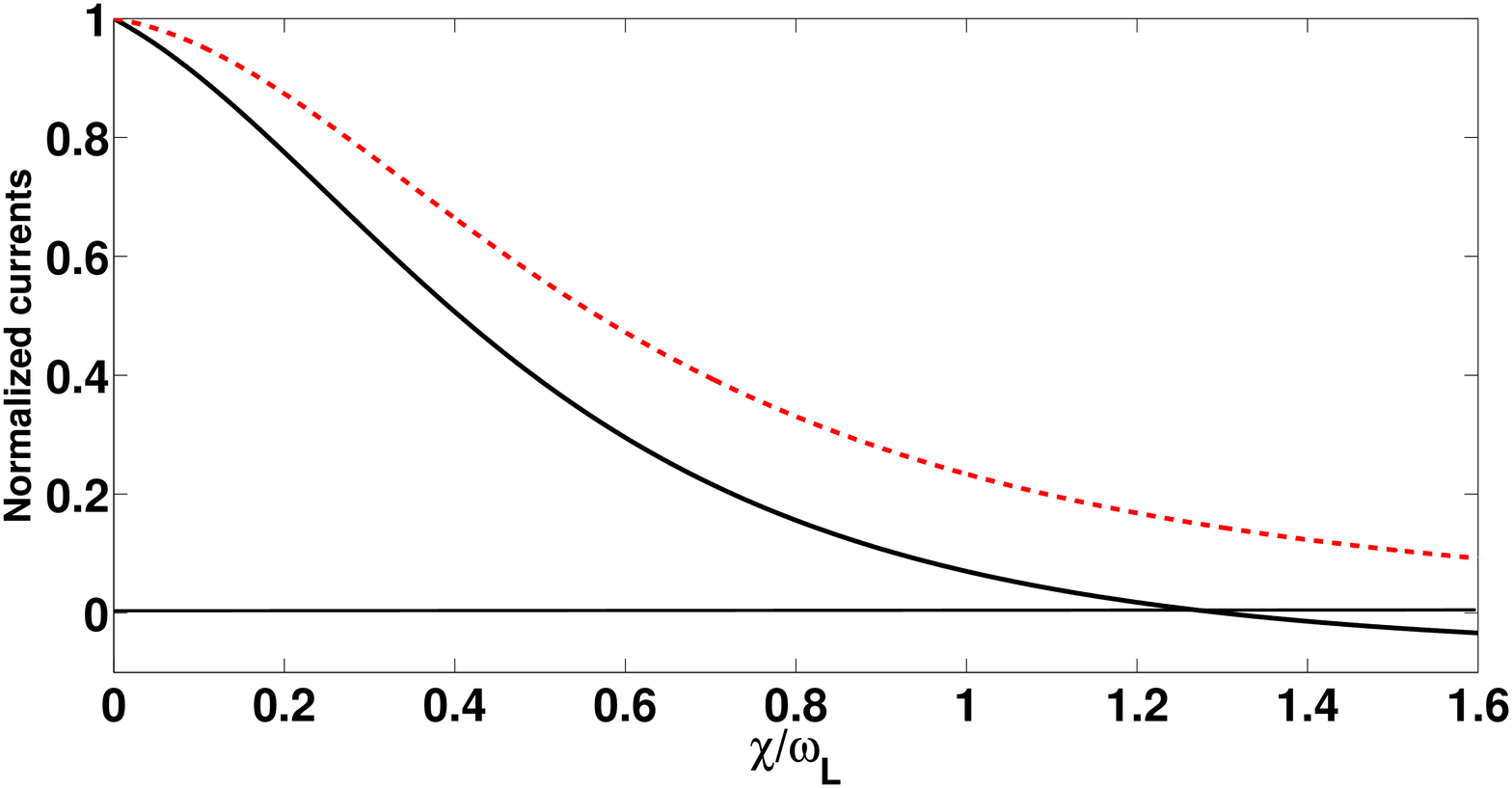}
\caption{Normalized forward current $I_{f}$ (continuous line) and reverse current $I_{r}$ (dashed line) as a function of $\chi/\omega_L$ for $\Gamma_R/\Gamma_L=0.3$. Currents are normalized with their respective values at $\chi=0$. The system-reservoir parameters are $J/\omega_L=0.05, \bar{n}_L-\bar{n}_R=0.5, \omega_R/\omega_L=1$ and $\langle \sigma_z\rangle=-1$. }
\label{JfJr}
\end{figure}

\begin{widetext} 
Current from the left reservoir $R_L$ to the right reservoir $R_R$ in the system shown in Fig. \ref{CoupledCavity} is called forward current. The expression for the forward current is
\begin{align}\label{forward}
I_{f}(\Delta n,\Gamma_L,\Gamma_R)=J^2\delta N\frac{(\omega_L\Gamma_R+\Gamma_L(\omega_R-\chi))}{(\chi^2-\Delta_c^2+\gamma^2)^2+4\gamma^2\Delta_c^2}((\Delta_c-\chi)^2+\gamma^2).
\end{align}\\

On exchanging ($\bar{n}_L, \Gamma_L)$ and $(\bar{n}_R,\Gamma_R)$, reverse current from the right reservoir $R_R$ to the left reservoir $R_L$ in the configuration given in Fig. \ref{CoupledCavityReverse} is
\begin{align}\label{reverse}
I_{r}(-\Delta n,\Gamma_R,\Gamma_L)=-J^2\delta N\frac{(\omega_L\Gamma_L+\Gamma_R(\omega_R-\chi))}{(\chi^2-\Delta_c^2+\gamma^2)^2+4\gamma^2\Delta_c^2}((\Delta_c-\chi)^2+\gamma^2).
\end{align}
The reverse current is taken as negative as the direction of flow is opposite to the forward current.
\end{widetext}
% If the system-reservoirs coupling are unequal and $\omega_L \neq |\omega_R-\chi|$,  $|I_{Lf}|\neq |I_{Lr}|$ as can be seen from Eqn. \ref{forward} and \ref{reverse}. Hence, the thermal current depends on the direction of flow through the system,exhibiting thermal rectification.
  The currents $I_{f}$ and $I_{r}$, normalized with their corresponding values for $\chi=0$ and $\Delta_c=0$, are shown as a function of the atom-field coupling strength $\chi$ in Fig. \ref{JfJr}. For non-zero $\chi$,  the magnitudes of the forward and reverse currents are different. Therefore, the system shows thermal rectification.  Importantly, if the parameters satisfy the condition given in Eqn. \ref{LowtoHigh}, the forward current changes the sign. As a result, $I_{f}$ and $I_{r}$ flow in same direction.\\    

Thermal rectification is quantified by rectification coefficient $R$ defined as 
\begin{align}\label{DefRectification}
R=-\frac{I_{f}}{I_{r}}.
\end{align}
If $R=1$, there is no rectification. For the system under consideration
\begin{align}\label{SystemRect}
R=\frac{\omega_L\Gamma_R+\Gamma_L(\omega_R-\chi)}{\omega_L\Gamma_L+\Gamma_R(\omega_R-\chi)}.
\end{align}
\begin{figure}
\centering
\includegraphics[width=9cm,height=6cm]{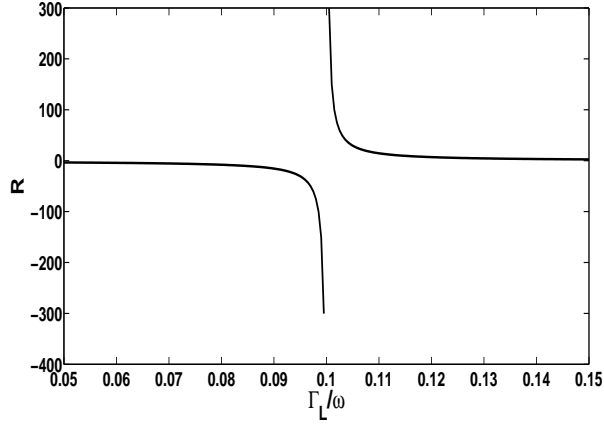}
\caption{Rectification $R$ as a function of $\Gamma_L/\omega$ for $\Gamma_R/\omega=0.2$. Here $\chi/\omega=1.5$ and $\langle \sigma_z\rangle=-1$. }
\label{Rectification}
\end{figure}
If $\Gamma_L=\Gamma_R$ or $\omega_L=|\omega_R-\chi|$, rectification coefficient $R$ becomes unity.  
Rectification coefficient $R$ is shown as a function of $\Gamma_L/\omega$ in Fig. \ref{Rectification} for the resonant case ($\Delta_c=0$).  Rectification is positive, zero, or negative depending on the parameters.  The system shows large rectification if 
\begin{align}\label{ZeroReverseCurrent}
\frac{\Gamma_L}{\Gamma_R}=\frac{\chi-\omega_R}{\omega_L},
\end{align} 
as seen in Fig. \ref{Rectification}. This originates  from the fact that the atom completely blocks the current in one direction (thermally insulating) and allows in the other direction (thermally conducting). Even though the system size is finite, rectification becomes infinity theoretically. %This observation is in contrast to the observation made in Ref. \cite{Landi}, wherein the rectification is finite in thermodynamic limit.
If $\Gamma_L/\omega$ is increases from values less than that satisfying Eqn. \ref{ZeroReverseCurrent} to higher values, $R$ jumps from negative value of large magnitude  to large positive value. Thus $R$ is very sensitive to changes in the parameters in that region.   Asymmetry can also be introduced with non-resonant cavities without an atom $(\chi=0)$ in any of the cavities. However, as seen from Eqn. \ref{SystemRect}, large rectification is not possible.
\section{Generalization to $N$-cavities}\label{NcavitySection}
It would be interesting to study the steady state heat transfer in $N$ coupled cavities containing a two-level atom in one of the cavities.  The Hamiltonian for the system is
\begin{align}\label{Ncavity}
\tilde{H}=\frac{\omega_0}{2}\sigma_z+\omega \sum_{j=1}^N a_j^\dagger a_j & +J\sum_{j=1}^{N-1} (a_j^\dagger a_{j+1}+a_j a_{j+1}^\dagger) \nonumber\\
                    & +\chi(\sigma_+\sigma_-+a^\dagger_m a_m\sigma_z).
\end{align}
The atom is embedded in the $m$th cavity and dispersively interacts with the cavity-field. The right most and the left most cavities in the array are coupled with two reservoirs $R_L$ and $R_R$ respectively. The density matrix $\tilde{\rho}$ of the system obeys
\begin{align}\label{MasterN}
\frac{\partial \tilde{\rho}}{\partial t}=-i[\tilde{H},\tilde{\rho}]+\mathcal{D}_L(\tilde{\rho})+\mathcal{D}_R(\tilde{\rho}),
\end{align}
where

\begin{align*}
\mathcal{D}_L(\tilde{\rho})=&\frac{\Gamma_L(\bar{n}_L+1)}{2}(2a_1\tilde{\rho} a_1^\dagger-a_1^\dagger a_1 \tilde{\rho}-\tilde{\rho} a_1^\dagger a_1)\nonumber\\
&+\frac{\Gamma_L \bar{n}_L}{2}(2a_1^\dagger\tilde{\rho} a_1-a_1 a_1^\dagger \tilde{\rho}-\tilde{\rho} a_1 a_1^\dagger),\\
\mathcal{D}_R(\tilde{\rho})=&\frac{\Gamma_R(\bar{n}_R+1)}{2}(2a_N\tilde{\rho} a_N^\dagger-a_N^\dagger a_N \tilde{\rho}-\tilde{\rho} a_N^\dagger a_N)\nonumber\\
&+\frac{\Gamma_R \bar{n}_R}{2}(2a_N^\dagger\tilde{\rho} a_N-a_N a_N^\dagger \tilde{\rho}-\tilde{\rho} a_N a_N^\dagger),
\end{align*}
 Here $\bar{n}_L$ and $\bar{n}_R$ are the mean number of photons in the reservoirs $R_L$ and $R_R$ respectively. Without loss of generality, we assume $\bar{n}_L > \bar{n}_R$.\\

Using Eqn. \ref{MasterN}, the equation of motion is
\begin{align}
\frac{d \langle G\rangle}{dt}=\frac{d}{dt} \text{Tr}(\tilde{\rho} G)=i[M_1,\langle G\rangle]+\{M_2,\langle G\rangle\}+M_3,
\end{align} 
where $\langle G\rangle=\langle A^\dagger A\rangle$ is the matrix whose elements are the expectation values of the operator elements of $A^\dagger A$. Here
\begin{align*}
A&=\text{Row}(a_1,a_2,...a_N,a_1\sigma_z,...,a_N\sigma_z),\\
A^\dagger&=\text{Column}(a_1^\dagger,a_2^\dagger,...a_N^\dagger,a_1^\dagger\sigma_z,...,a_N^\dagger\sigma_z).
\end{align*}
Further  $[M_1,\langle G\rangle]=M_1\langle G\rangle-\langle G\rangle M_1$ and $\{M_2,\langle G\rangle\}=M_2\langle G\rangle+\langle G\rangle M_2$. 
 The transformation matrices are
\begin{align*}
 M_1=&I_{2\times 2} \otimes H_c+\sigma_x \otimes X,\\
 M_2=&I_{2\times 2} \otimes \text{Diag}\left(-\frac{1}{2}\Gamma_L,0,...0,-\frac{1}{2}\Gamma_R\right)_{N\times N},\\
 M_3=&I_{2\times 2} \otimes \text{Diag}\left(\Gamma_L \bar{n}_L,0,...0,\Gamma_R\bar{n}_R\right)_{N\times N}\\
   &~~~~~~+ \sigma_x \otimes \text{Diag}(\Gamma_L \bar{n}_L\langle\sigma_z\rangle,0,...0,\Gamma_R\bar{n}_R\langle \sigma_z\rangle)_{N\times N},
\end{align*}
where $I_{2\times 2}$ is the identity matrix of dimension $2$ and $\sigma_x$ is the Pauli matrix. The matrix 
\[
H_c=\left(\begin{matrix}
\omega &  J & 0  & \ldots & 0\\
J &  \omega & J& \ldots & 0\\
\vdots & \vdots & \ddots & \vdots\\
0  &   0       &\ldots &\omega & J\\
0  &   0       &\ldots &J & \omega
\end{matrix}\right)_{N\times N},
\]
and the matrix elements of $(X)_{N\times N}$ are zero except $X_{m,m}=\chi$.\\
\begin{figure}[h!]
\centering
\includegraphics[width=9cm,height=6.2cm]{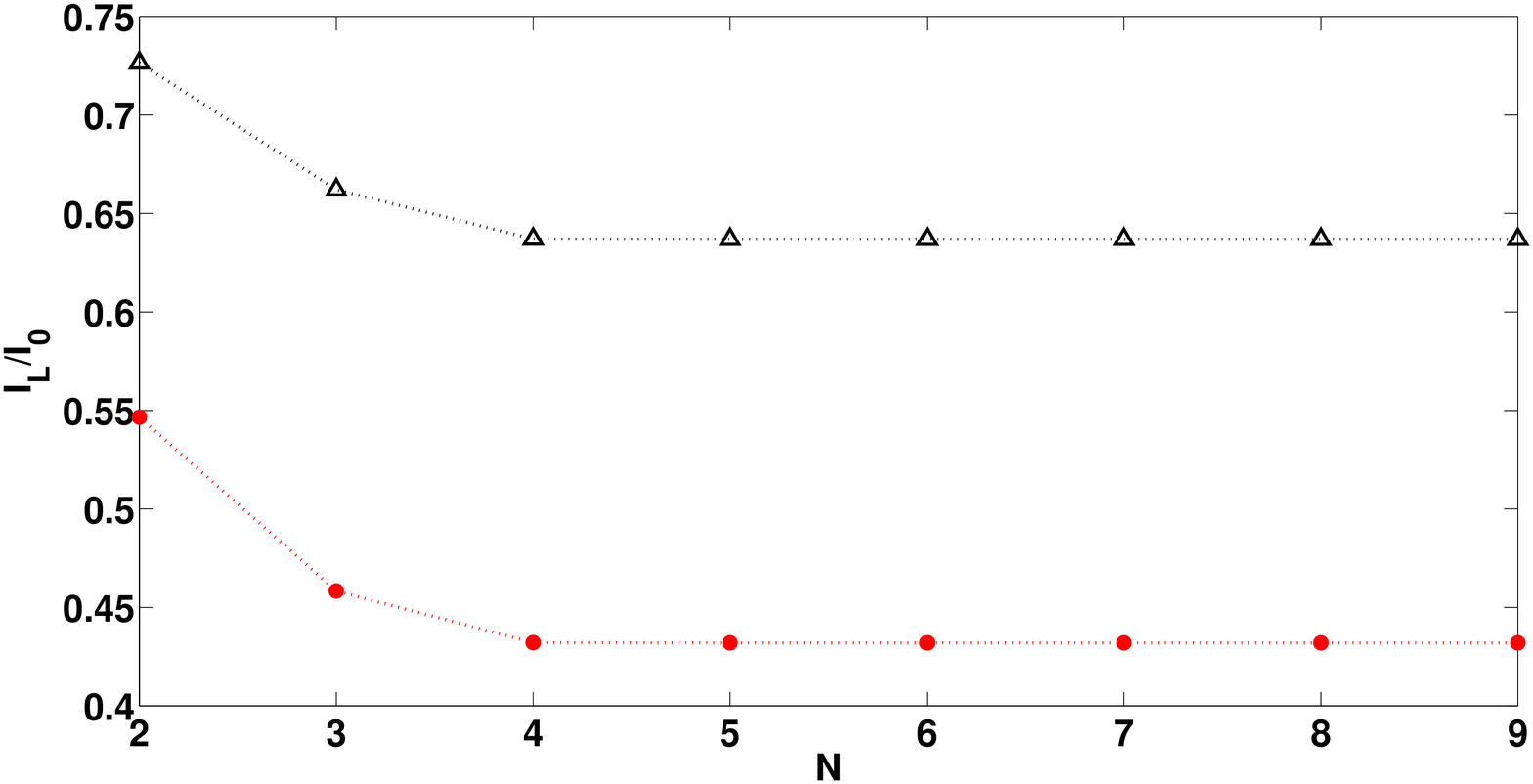}
\caption{Ratio of currents $I_L/I_0$ as a function of $N$ for $\chi/\omega=0.15$ (circle) and $0.1$ (triangle). The atom is embedded in the last cavity ($m=N$). Here $J/\omega=0.05, \bar{n}_L-\bar{n}_R=0.5, \langle\sigma_z\rangle=-1, \Gamma_L/\omega=\Gamma_R/\omega=0.15$. }
\label{Sizedependent}
\end{figure}

Using $\tilde{H}$ given in Eqn. \ref{Ncavity} in the continuity equation (refer Eqn. \ref{Cont}), the current in the system is 
\begin{align}\label{CurrentNcavities}
I_L=\Gamma_L\left[(\bar{n}_L- \right. &\langle a_1^\dagger a_1\rangle_{ss})(\omega+\chi\langle \sigma_z \rangle\delta_{m,1})\nonumber\\
&\left.-\frac{J}{2}(\langle a_1^\dagger a_2\rangle_{ss}+\langle a_1 a_2^\dagger \rangle_{ss})\right].
\end{align} 
Here $\delta_{m,1}$ is Kronecker delta.
If there is no atom in the array, the coherence term $\langle a_j^\dagger a_{j+1}\rangle$ is purely imaginary \cite{Asad}. The contribution of the coherence term to the current vanishes as $I_{coh}=\frac{J}{2}(\langle a_1^\dagger a_2\rangle_{ss}+\langle a_1 a_2^\dagger \rangle_{ss})=0$. Consequently, current in the cavity array is
\begin{align}\label{AbsentAtom}
I_L(\chi=0)=I_0=\frac{4\omega J^2\Gamma_L\Gamma_R}{(4J^2+\Gamma_L\Gamma_R)(\Gamma_L+\Gamma_R)}(\bar{n}_L-\bar{n}_R).
\end{align}
Note that the current $I_0$ is independent of the size of the array, in violation of Fourier's law. This feature is similar to the system-size independent current in the case of ballistic transport \cite{Zurcher, Gaul, Asad}. This comparison indicates that the mean free path of the photons scales in proportion to the number of cavities $N$.\\

\begin{figure}[h!]
\centering
\includegraphics[width=9cm,height=5cm]{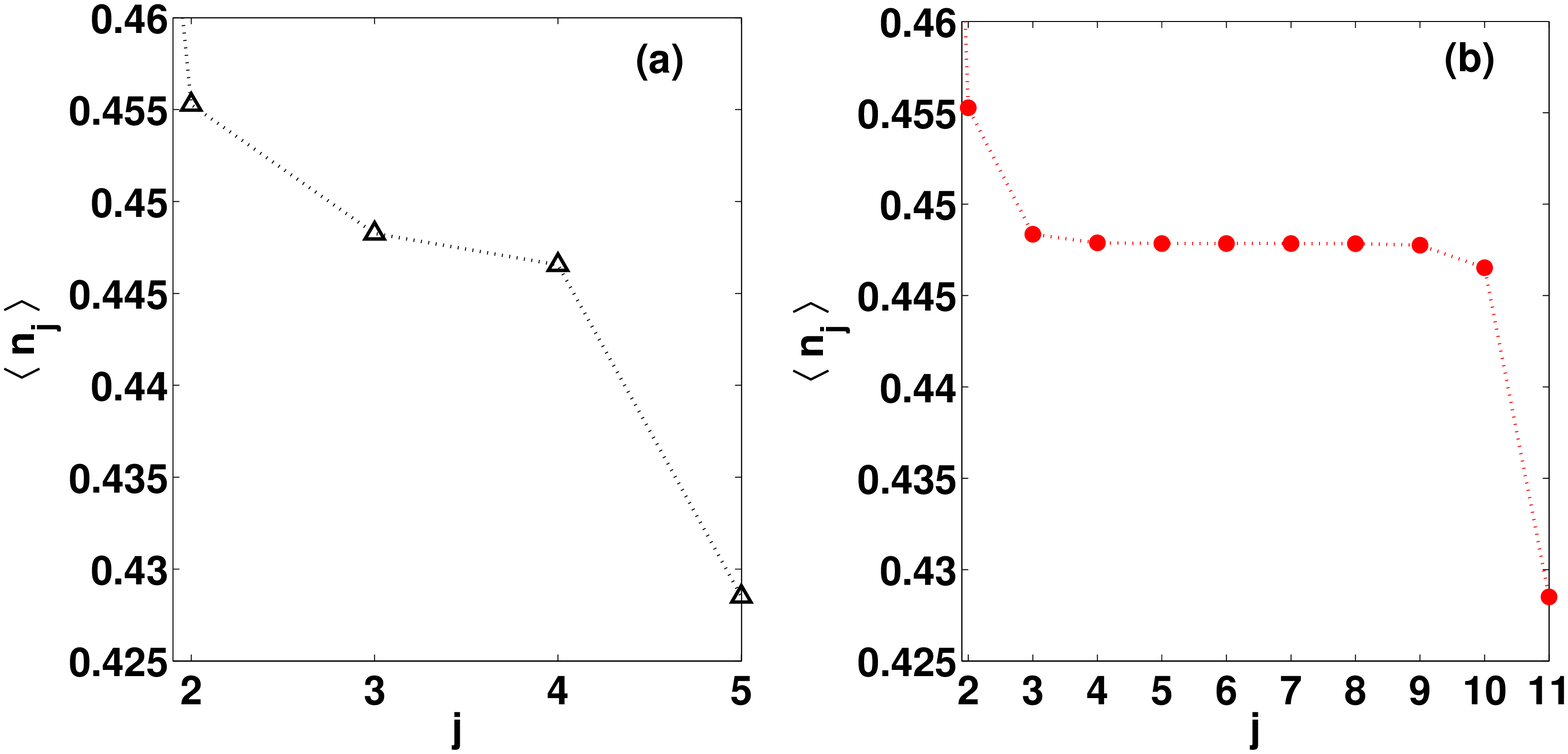}
\caption{Steady state mean photon number in the intermediate cavities for arrays of length $(a)N=6$ and $(b)N=12$. The atom-field coupling strength is chosen to be $\chi/\omega= 0.1$.   Here $J/\omega=0.05, \bar{n}_L-\bar{n}_R=0.5, \langle\sigma_z\rangle=-1, \Gamma_L/\omega=\Gamma_R/\omega=0.15$.}
\label{temperatureprofile}
\end{figure}
Mean free path is different from the array size if an atom is embedded in one of the cavities. The atom is considered to be in the last cavity of the array, \textit{i.e.}, $m=N$, to keep the mean free path as close to the size of the array. This helps to understand the emergence of diffusive character if there is a single scatterer. The normalized current $I_L/I_0$ as a function of size of the array $N$ is shown in Fig. \ref{Sizedependent} for a fixed temperature difference $(\bar{n}_L-\bar{n}_R)$. It is to be noted that by increasing the size of the array, the steady state current significantly decreases and asymptotically approaches a constant value. The current is size dependent for smaller array. Thus the atom is able to introduce diffusive character. It saturates with further increase in size and becomes nearly size independent, which is at odds with the  Fourier's law. Thus,  the transport is of ballistic type.  If many cavities in the array contain atoms, the heat transport may be expected to be diffusive.  This is plausible as the effect of dephasing by all the atoms effectively reduces the mean free path for the photons \cite{Brune, Asad}. \\   

The transition from diffusive to ballistic transport as size of the array increases, can be understood by calculating the mean photon numbers $\langle n_j\rangle=\langle a_j^\dagger a_j\rangle$ (known as local temperature \cite{Asad}) of the respective  cavities in the array. The steady state mean photon number $\langle n_j\rangle$ in the intermediate cavities for arrays containing $6$ and $12$ cavities are shown in Fig. \ref{temperatureprofile}$(a)$ and $(b)$ respectively. Gradient in the mean photon number is noticed in Fig. \ref{temperatureprofile}$(a)$. This implies that the transport is diffusive \cite{Reider, Hu}. For larger size array, for instance $N=12$, the gradient in mean photon number approaches zero and the current is independent of the system size.  Essentially, the change in mean free path in the presence of a scatterer at the end of the array is insignificant for large array.  Consequently,  the photon transport is not diffusive.   
%%%%%%%%%%%%%%%%%%%%%%%%%%%%%%%%%%%%%%%%%%%%%%%%%%%%%%%%
\section{Summary}\label{Summary}
Mesoscopic systems offer interesting possibilities when it comes to thermal properties. A system of two coupled cavities connected between thermal reservoirs provides a conduit for heat flow between the reservoirs. If a dispersively interacting atom is placed in one of the cavities, thereby providing an asymmetry in the system, many of the thermal transport properties can be tailored. The present system switches from a thermally insulating state to a conducting one, depending on whether the atom is in its ground state or excited state. If the atomic state changes from the excited state to the ground state, current through the system becomes zero or reversed depending on the system-reservoir coupling strengths and the cavity frequencies. The reversal of current implies that the effective thermal conductivity is negative.\\ 

 The presence of the atom changes the magnitude of the current on exchange of the reservoirs along with the coupling strengths, which leads to thermal rectification. Large rectification is possible if the parameters are chosen to make the system thermally resistive either for the forward current or the reverse current.\\
 
%In an array of $N$ linearly coupled resonant cavities, current is independent of the array size, which is characteristic of ballistic transport.

If a cavity  array contains a two-level atom in one of the cavities, the magnitude of current depends on the number of cavities. This size-dependence indicates that the thermal current through the array is analogous to the diffusive heat transport. If there is only a single atom in a large array, it is not possible to completely recover the diffusive transport. Single atom does not provide enough dephasing to recover the diffusive character.\\\\
%%%%%%%%%%%%%%%%%%%%%%%%%%%%%%%%%%%%%%%%%%%%%%%%%%%%%%%%%%%%
\textbf{Data accessibilities}\\
This paper has no data.\\
\textbf{Competing interest}\\
We have no competing interest.\\
\textbf{Authors' contribution}\\
Both the authors formulated and analyzed the problem. Both contributed to the interpretation of the results.\\
\textbf{Funding statement}\\
There is no funding.\\
\textbf{Ethics statement}\\
It does not apply.  
%%%%%%%%%%%%%%%%%%%%%%%%%%%%%%%%%%%%%%%%%%%%%%%%%%%%%%%%%%%

%\bibliographystyle{apsrev4-1}
%\bibliography{TheramswitchphotoniccurrentNotes}

\end{document}